￼
\documentclass[twocolumn,a4]{article}
    \title{{\bf Generation High resolution 3D model from natural language by Generative Adversarial Network}} 
    \author{Kentaro Fukamizu, Masaaki Kondo, Ryuichi Sakamoto}
    \date{}
    
    \usepackage[top=20truemm,bottom=20truemm,left=15truemm,right=15truemm]{geometry}
    \usepackage{amsmath,amssymb}
    \usepackage{amsfonts}
    \usepackage[pdftex]{graphicx}
    \usepackage{here}
    \usepackage{ascmac}
    \usepackage{bm}
    \usepackage{listings}
    \usepackage{subfig}

\begin{document}

\maketitle

\begin{abstract}
    {\bf
    Since creating 3D models with 3D designing tools is a heavy task for human, there is a need to
    generate high quality 3D shapes quickly form text descriptions. In this paper, we propose a
    method of generating high resolution 3D shapes from natural language descriptions.  Our method is
    based on prior work \cite{chen2018text2shape} where relatively low resolution shapes are
    generated by neural networks. Generating high resolution 3D shapes is difficult because of the
    restriction of the memory size of a GPU or the time required for neural network training. To
    overcome this challenge, we propose a neural network model with two steps; first a low resolution
    shape which roughly reflects a given text is generated, and second corresponding high resolution
    shape which reflects the detail of the text is generated. To generate high resolution 3D shapes,
    we use the framework of Conditional Wasserstein GAN. we perform quantitative evaluation with
    several numerical metrics for generated 3D models. We found that the proposed method can improve
    both the quality of shapes and their faithfulness to the text.
    }
    \end{abstract}

\section{Introduction}
People usually use their natural languages to communicate their thoughts and emotions to others.
Recently, artificial intelligence (AI) technology allows us to use these natural languages for
operating and controlling the computer systems. For example, it is becoming possible for an AI
engine to generate 3D shapes from descriptions of a natural language.

When creating 3D models, 3D designers commonly rely on expensive modeling software, such as
Maya, Blender, and 3DSMAX, and they need to spend very long time to create satisfied quality of
3D models with these tools. Even after becoming an expert of these tools, it takes a long time
to create one 3D model. Compared with using 3D modeling software, if human already has an image
of the target object in her/his brain, it is very easy to express the outline of the shapes in
the form of text descriptions. If an AI engine can generate 3D shapes quickly form text
descriptions, it is possible to reduce the time taken for these heavy tasks.

There have been research efforts to generate 3D models from vectors of encoded text descriptions
by using Generative Adversarial Network(GAN)\cite{GAN}.
For example, in \cite{VoxNet}, it is possible to generate 3D shapes with rough categories, but
it is not possible to generate the fine models of them with colors or details of each part.

Prior work shown in \cite{chen2018text2shape} successfully generates 3D
shapes with colors from text descriptions. However the resolution of the shapes is still low.
In the work, text is first converted to a vector representation via an encoder.
Next, a corresponding 3D shape is generated by the Generator from the vector.  This Generator is
based on a deep-learning methodology and a GAN framework. When the vectorized text is input to
the Generator, it is combined with noises to ensure the flexibility of the Generator.

The output of the 3D shapes is created through 3D deconvolution operations and then input to a
Critic network.  The Wasserstein distance between the actual 3D shape and the generated one is
calculated.  On one hand, the Critic network attempts to calculate the Wasserstein distance
accurately. On the other hand, the Generator attempts to minimize the distance calculated by
Critic. This ensures that the probability distribution of the actual 3D shape and the generated
one becomes closer. The details of the algorithm will be described later in Section ??.

In the previous work, the output is generated in the form of 3D voxels. As for voxel generation,
learning was done with voxel size of $(x, y, z, color) = (32, 32, 32, 4)$.  Since, in general,
3D model designers do not create 3D model in the form of voxels but in the form of mesh
structure, voxels can be converted to mesh with the some methods such as Marching
Cube\cite{Marching}. However, if the resolution of the voxels is low, the appearance of the
transformed mesh becomes considerably coarse. In addition, it becomes difficult to display
sufficient details written in the text with low resolution voxels. Therefore, in this research,
we propose another GAN learning methodology to generate voxels with higher resolution.
Specifically, we use voxel size of $(x, y, z, color) = (64, 64, 64, 4)$.

One of the ways to make the resolution higher is using larger neural network models. However,
this cause tremendous increase in the number of parameters required to learn.
If we simply use this large neural network model with a number of parameters,
learning time increases significantly and sometimes learning is unstable.
Therefore, in our proposed methodology, we reconsider the role of the Critic network
of the previous research so that learning phase can be fast and stable.

Throughout this paper, we propose several GAN models where the rolse of the Critic network differ,
for example, whether or not focusing on preciseness of the text descriptions.  Since these
models have advantages and disadvantages for various indices, we also introduce several metrics
to compare the effectiveness of these proposed models.

The rest of this paper is organized as follows. In the next section, ..... 

\section{Related Work}

\subsection{Generative Adversarial Network (GAN)\cite{GAN}}

In the Generative Adversarial Network (GAN) framework, two networks called Generator and
Discriminator are learned. Generator is trained to generate the data similar to the training data from
latent vector. Discriminator learns to discriminate between the training data and the
data generated by Generator. Generator and Discriminator is trained alternately with this
mechanism, and finally it is expected that Generator can make the data similar to the training
one.

These processes can be expressed by the following mathematical expression,
\begin{align*}
    \min_{G}\max_{D}V(D,G)&=\mathbb{E}_{x\sim p_{data}(x)}[\log D(x)]\\
    &\quad+\mathbb{E}_{z\sim p_{z}(z)}[\log (1-D(G(z)))]
\end{align*}
where, $G$, $D$, and $x$ are Generator, Discriminator, and training data, respectively. This
expression shows the objective function of $D,G$. Here, we use $z$ as a noise dor the latent
vector and $G$ generates data from the noise $z$. $D(x)$ means the probability
that $x$ is regarded as the training data.

\subsection{Conditional GAN (CGAN)\cite{CGAN}}

In the original GAN, all elements of the latent vector which input to $G$ are noise based on certain probability distribution.
However, if noise is taken as an input, it is difficult to specify and generate the data which you want to generate specially.
On the other hand, CGAN can express what we want to generate, by inputting a label as latent vector into Generator
and Discriminator. In the CGAN, the latent vector of what you want to generate is encoded
(from image, text. etc.) before generation.

In CGAN, the latent vector $l$ which is appropriately generated by the training data is combined
with the noise vector $z$ and input to the Generator. The reason for combining with the noise
vector is that this ensures Generator having diversity. Otherwise, it may
limit its possible output, determined only by the latent vector.
In addition to that, this helps make the output more robust even for some parts
where latent vector cannot describe sufficiently (such as details of the background of an image).

It is common in usual GAN frameworks to use combined latent vector for an input to Generator
so that the Discriminator recognizes the generated data as training data.
The difference of CGAN is that it uses the combined vector $l$
as an input to Discriminator. By doing so, Discriminator can judge whether the output
from the Generator is actually linked with the corresponding latent vector description.

To implement the above point, the objective function of CGAN is expressed as follows:
\begin{align*}
    \min_{G}\max_{D}V(D,G)&=\mathbb{E}_{x\sim p_{data}(x)}[\log D(x)]\\
    &\quad+\mathbb{E}_{z\sim p_{z}(z), l\sim p_{l}(l)}[\log (1-D(G(l,z)))]
\end{align*}

In order to strengthen the capability of identifying whether output is generated with
the corresponding latent vector, the literature \cite{AttnGAN} proposes a way to train
the data so that $D$ should output $0$ for training data that differs with the
description of the latent vector. In this case, the objective function is expressed as follows:
\begin{align*}
    \min_{G}\max_{D}V(D,G)&=\mathbb{E}_{x\sim p_{data}(x)}[\log D(x)]\\
    &\quad+\mathbb{E}_{x\sim p_{mis}(x)}[\log (1-D(x))]\\
    &\quad+\mathbb{E}_{z\sim p_{z}(z), l\sim p_{l}(l)}[\log (1-D(G(l,z)))]
\end{align*}
where $p_{mis}(x)$ is the probability distribution of the training data that mismatched with
the latent vector.

\subsection{WassersteinGAN(WGAN)\cite{WGAN}}


The aim of WassersteinGAN (WGAN) is to bring the probability distribution of the generated
voxel, $\mathbb{P}_g(x)$, close to the probability distribution of the training data,
$\mathbb{P}_r(x) $. The simplest way of doing this is minimizing Kullback-Leibler (KL) divergence
which is one of the metrics to observe distances between two probability distributions.
KL divergence ($KL$) between the probability distribution $ \mathbb{P}_r $ and $ \mathbb{P}_g $ is
calculated as follows:
\begin{align*}
    KL(\mathbb{P}_r||\mathbb{P}_g)=\int \log\left(\cfrac{P_r(x)}{P_g(x)}\right)P_r(x)dx
\end{align*}
In the case of GANs, it is not possible to numerically calculate $KL$, and hence compute the loss
function for them, because a specific probability distribution is not assumed in GANs.

Instead of using $KL$ directly, the GAN try to minimize Jensen-Shanon divergence (JSD).
Although $KL$ is asymmetrical, JSD can be symmetrical. The mathematical expression of JSD is
represented as follows:
\begin{align*}
    \mathbb{P}_A&=\cfrac{\mathbb{P}_r+\mathbb{P}_g}{2}\\
    JSD(\mathbb{P}_r||\mathbb{P}_g)&=\cfrac{1}{2}KL(\mathbb{P}_r||\mathbb{P}_A)+\cfrac{1}{2}KL(\mathbb{P}_g||\mathbb{P}_A)
\end{align*}

As for loss of the GAN shown below;
\begin{align*}
    V(D,G)&=\mathbb{E}_{x\sim p_{data}(x)}[\log D(x)]\\
    &\quad+\mathbb{E}_{z\sim p_{z}(z)}[\log (1-D(G(z)))]
\end{align*}
obviously JSD becomes maximum in the case of
\begin{align*}
    D^*(x)=\cfrac{P_r(x)}{P_r(x)+P_g(x)}
\end{align*}

In this condition, the objective function becomes as follows
\begin{align*}
    V(D^*,G)=2JSD(\mathbb{P}_r||\mathbb{P}_g)-2\log 2
\end{align*}
In other words, while $D$ accurately approximates JSD, $G$ learns to minimize JSD which is
calculated by $D$. However, in training GANs, it is very important to carefully adjust
learning balance between Discriminator and Generator or learning rate of them. If the
Discriminator's training is insufficient, the Generator will minimize the incorrect JSD. On
the other hands, if Discriminator's training is too enough, the gradient for the parameters
of Discriminator will be small, making Generator's training infeasible.

As discussed, the success of learning of GANs depends on learning parameters. Specially when
learning a model with a large number of parameters such as for 3D voxel creation in this research,
adjustment of those learning parameters is extremely difficult. To overcome this challenge,
One of prior work introduces another index to measure the distance between probability distributions
instead of using JSD. In WassersteinGAN\cite{WGAN}, Wasserstein distance is introduced as a
metric of distance. Wasserstein distance is expressed as follows:
\begin{align*}
    W(\mathbb{P}_r,\mathbb{P}_g)=\inf_{\gamma\in\prod(\mathbb{P}_r,\mathbb{P}_g)}\mathbb{E}_{(x,y)\sim\gamma}[||x-y||]
\end{align*}
where $\prod(\mathbb{P}_r, \mathbb{P}_g)$ denotes the set of all joint distributions
$\gamma(x,y)$ whose marginals are $\mathbb {P}_r$ and $\mathbb {P}_g$, respectively.
Intuitively, $\gamma(x, y)$ indicates how much ``mass'' must be transported from $x$ to $y$
for converting the probability distributions $\mathbb{P}_r$ into the probability distribution
$\mathbb{P}_g$.  Originally, this is a metric used for optimal transport problems. This makes
measuring the distance between low dimensional manifolds possible.

The Wasserstein distance can a better way to describe the distance than JSD, but it is difficult
to calculate.  According to Kantorovich-Rubinstein duality \cite{soutui}, it can be expressed
using the 1-Lipschitz function $f$ as follows:
\begin{align*}
    W(\mathbb{P}_r,\mathbb{P}_g)=\sup_{||f||_L\leq 1}\mathbb{E}_{x\sim P_r}[f(x)]-\mathbb{E}_{x\sim P_g}[f(x)]
\end{align*}
where $ x\in\chi$ and $f:\chi\rightarrow \mathbb{R}$. The meaning of 1-Lipschitz function is that
the slope of a straight line of arbitrary $x,x'\in \chi $ does not exceed $1$.  Here, if
$f$ is a function $\{f_w\}_{w\in W}$ represented by some parameters and $x \sim P_g$ follows
$ g_\theta (z): \mathbb{R}^n \rightarrow \chi$, we can express $ W(\mathbb{P}_r, \mathbb{P}_g)$
as follows:
\begin{align*}
    W(\mathbb{P}_r,\mathbb{P}_g)=\max_{w\in W}\mathbb{E}_{x\sim P_r}[f_w(x)]-\mathbb{E}_{z\sim P_z}[f_w(g_\theta(z))]
\end{align*}
In order to satisfy the 1-Lipschitz condition, it suffices that each parameter fit into the
compact space, that is, the absolute value of the weight parameter $ w $ is clipped to a
certain value $ c $.  Discriminator is called Critic to distinguish it from the original
GAN.

In WGAN, the Critic and the Generator networks learn alternately until Wasserstein distance
converges.  While the Critic attempts to calculate the Wasserstein distance between training
data and generated data accurately, the Generator attempts to bring the probability
distribution of generated data close to that of the training data by minimizing the
Wasserstein distance calculated. Given that gradients do not disappear even if the Wasserstein
distance converges completely WGAN is very stable in learning.  Therefore adjustment of the
learning balance between the Critic and the Generator is unnecessary.

In conventional GANs, since the Critic and the Generator networks use different loss
functions, the loss values do not converge even if training fairly progress. The problem here
is that the timing of finishing training is difficult to find out.  However, the loss value of
the Critic always goes toward converge in WGAN.  Therfore, one of the advantages of WGAN is
that decrease of the loss value always correlates with improvement in the quality of the
generated data.

However, one of the disadvantage in WGAN is clipping weight parameters. If the weight
parameters are clipped, the weights become polarized to the clipped boundary values, resulting
in the gradient explosion or disappearance. This cause delay in learning. Here, the optimized
Critic has a characteristic that it has a slope whose norm is $1$ at almost all points below
$\mathbb{P}_r $ and $ \mathbb{P}_g $\cite{WGANGP}. Based on this feature, WGANgp is
proposed\cite{WGANGP}.  In WGANgp, a penalty term is introduced in the loss function so that
the Critic has a slope whose norm is $1$ at almost all points below $ \mathbb{P}_r $ and $
\mathbb{P}_g$. This allows it to be optimized without clipping the weights.
Let $D(x)$ be the output of the Critic. Then the loss is expressed as follows:
\begin{align*}
    L_{WGANGP}&=\mathbb{E}_{x\sim P_g}[D(x)]-\mathbb{E}_{x\sim P_r}[D(x)]\\
    &\quad+\lambda \mathbb{E}_{x'\sim P_{gp}}[(||\nabla_{x'}D(x')||_2-1)^2]
\end{align*}
where $x'=\epsilon x+(1-\epsilon)\hat{x}$, at $\epsilon\sim U[0,1]$, $x\sim P_r$, $\hat{x}\sim P_g$.
By providing the penalty to the gradient, the weights have diversified values without polarizing,
giving higher model performance.

\subsection{text2shapeGAN\cite{chen2018text2shape}}

As mentioned in Section 1, this paper is based on prior work which generates 3D voxels from
texts\cite{chen2018text2shape}. In text2shapeGAN (TSGAN), they generate 3D voxels stably
by combining CGAN and WGANgp techniques. In TSGAN, the Critic not only evaluates how realistic the
generated voxels look like, but also how faithfully the generated voxels reflect texts. The objective
function of TSGAN is as follows:
\begin{align*}
    L_{CWGAN}&=\mathbb{E}_{t\sim p_\tau}[D(t,G(t))]+\mathbb{E}_{(t,s)\sim p_{mis}}[D(t,s)]\\
    &\quad-2\mathbb{E}_{(t,s)\sim p_{mat}}[D(t,s)]+\lambda_{GP}L_{GP}\\
    L_{GP}&=\mathbb{E}_{(t,s)\sim p_{GP}}[(||\nabla_t D(t,s)||_2-1)^2\\
    &\qquad+(||\nabla_s D(t,s)||_2-1)^2]
\end{align*}
where $t$ is the text embedding, $s$ is the 3D voxel, and $p_\tau$ is the probability
distribution of the text embedding. In addition, $p_{mat}$ and $p_{mis}$ are
the probability distribution of matching text-voxel pairs and mismatching text-voxel pairs,
respectively. Note that they sum up gradients for all the input variables of $D$ to make
the gradient penalty.
\begin{figure}[htbp]
    \centering
    \includegraphics[width=7cm]{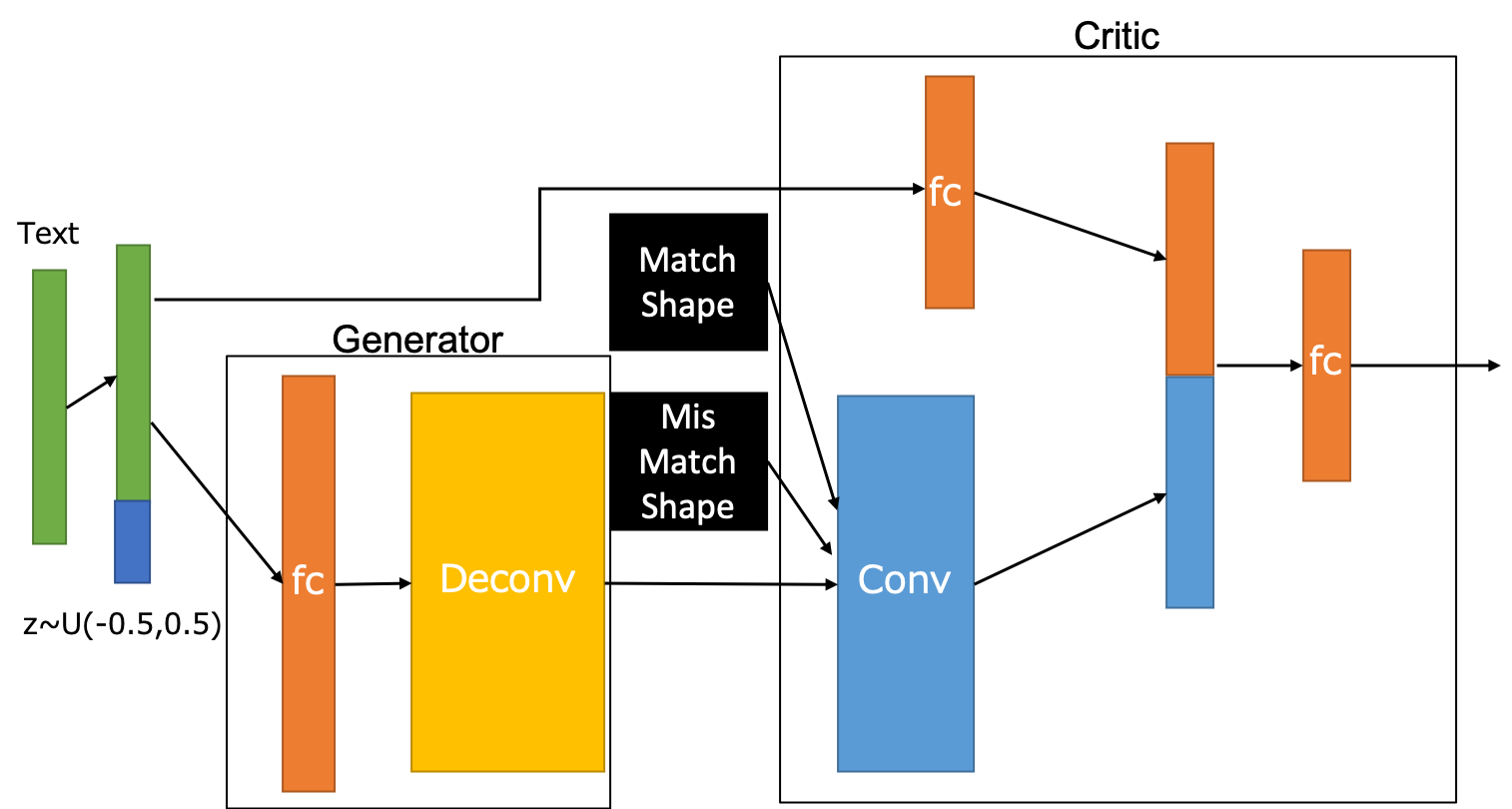}
    \caption{The model of TSGAN}
    \label{TSGAN}
\end{figure}

Fig.\ref{TSGAN} shows the model of TSGAN. First, it combines text embedding with a noise
vector. This is an input to the Generator and a set of 3D voxels is output by deconvolution.
As the last layer of the Generator, it has a sigmoid layer so that the output value is restricted
from $0$ to $1$. The generated 3D voxels are input to the Critic and it is transformed into a
one-dimensional vector via convolutional layers. It is combined with the text embedding and then
passed to the fully connected layers. The output of the final fully connected layer is a scaler value.
The output of the Generator is a set of voxels of $ (x, y, z, color) = (32, 32, 32, 4) $.

\section{Proposed Method}

\subsection{Approach}

As the simplest approach to generate high resolution 3D shapes, it is conceivable that we can
add a higher resolution deconvolution layer to the model of TSGAN. However, the number of
parameters to be added becomes too large to compute because it needs to deal with
three-dimensional data for learning. In this research, we assume to use a GPU for faster
learning speed, but the number of parameters that a GPU can store in its memory for learning
is limited, depending on the memory size of the GPU.

If the number of parameters is large, the problem is not only the fact that the learning is
sometimes terminated by the lack of memory, but also the number of epochs required for
training dramatically increases.  Since the training for three-dimensional data is
proportional to the order of cubic, training will not be finished in realistic time. Even
though we can limit the number of epochs, the generated voxels may be collapsed. For these
reasons, simply adding higher resolution layer does not work well.

To overcome the challenges described above, our approach is dividing the tasks into two steps;
one is for generating a low resolution shape which roughly reflects the target text (StageI)
and the other is for generating corresponding high resolution shape (StageII) by using the
knowledge of StackGAN\cite{StackGAN}. In StackGAN, StageI generates a low resolution image by
roughly deciding color distribution and placement of it. In StageII,
low resolution image generated in StageI is input to some convolution layers, then
combined with the text embedding, which is sent to the Residual layer. Finally,
a high resolution output image is generated via deconvolution layers.
In the proposed method, we use TSGAN as StageI and constructing a new model for StageII to
generate high resolution voxels. The following sections describe the details of these two stages.

\subsection{Low resolution task (StageI)}

\begin{figure}[htbp]
    \centering
    \includegraphics[width=7cm]{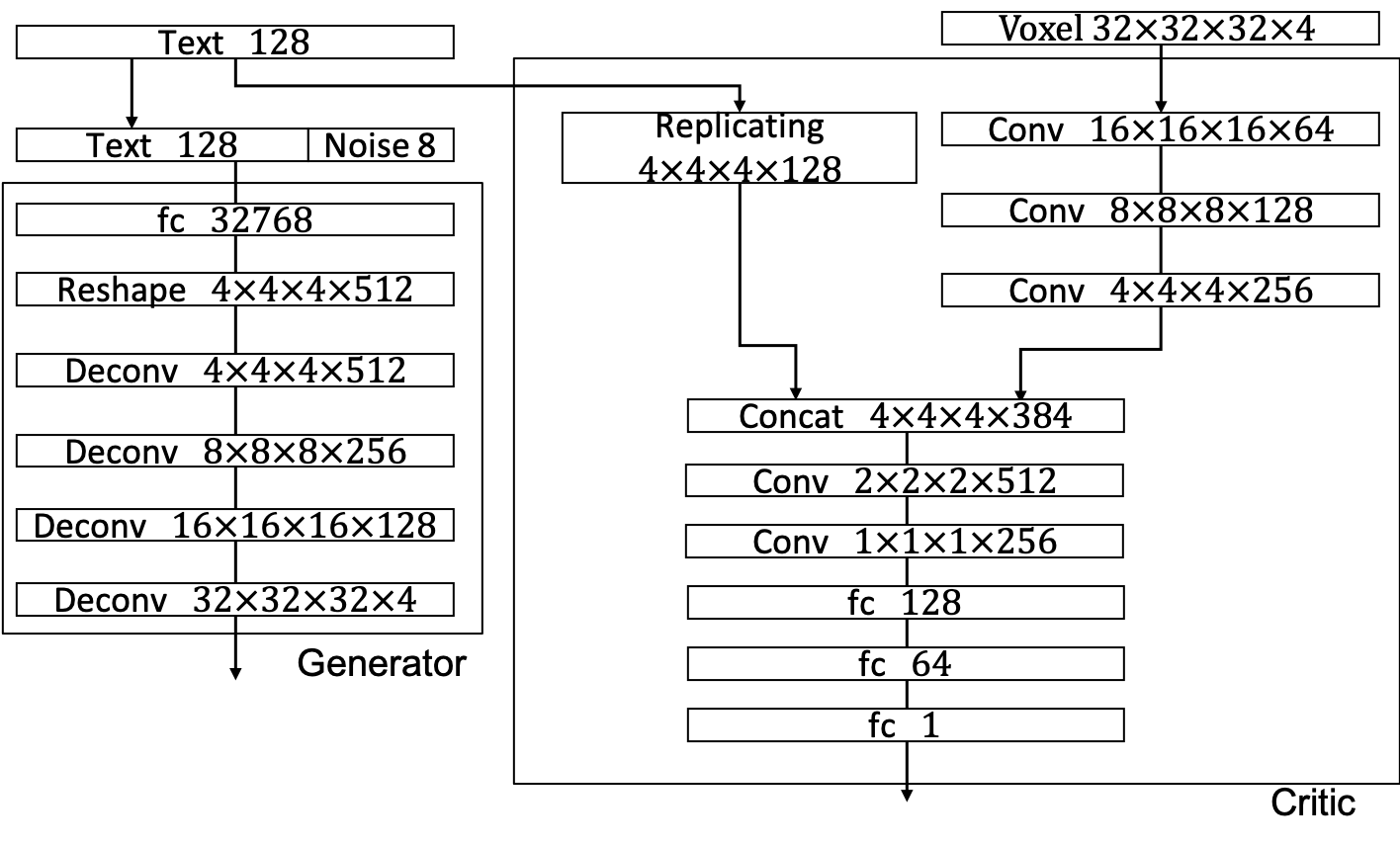}
    \caption{The network model of StageI}
    \label{stage1}
\end{figure}

The Generator tries to create rough shapes of resulting voxels at this stage. Unlike TSGAN,
StageI in this research does not need to generate voxels which are strictly faithful to the
input text since the details described in the text are reshaped at StageII. Instead of using
latent vectors combined with additional noises as proposed in \cite{chen2018text2shape}, we use only latent
vectors as the input to the Critic since we found that it can create sufficient level of
voxels and achieve faster convergence of training. In addition to that, we found a problem in TSGAN.
In TSGAN, generated voxels are first converted to one-dimensional vectors through convolutional and
fully connect layers of the Generator, and then combined with text latent vectors.
This causes some cases that spatial information of voxels is lost, resulting in
the lack of meaningful connection between the voxels and corresponding texts. Therefore, we
spatially duplicated the text embedding vectors and combined them with the convolved voxels
to hold the spatial feature as like StackGAN\cite{StackGAN}. Fig.\ref{stage1} shows our proposed
network model for StageI.

\subsection{High resolution task (StageII)}


In the high resolution task, the role of the Critic network varies depending on the type of
input variables or loss functions. In contrast with StageI which decides the rough color and
shape of voxels, we can consider two training models in StageII; focusing only on heightening
resolution or extending the existing training models by refining the prior work. Therefore,
we propose the following two models for StageII:

\begin{itemize}
\item (v0): This model supposes that the faithfulness to text is sufficient in StageI, so that
  binding to text is relaxed. The Critic network focus on whether the generated high
  resolution shapes are correct or not.
\item (v1): Like TSGAN, the Critic network focus on whether voxel is accurately generated from the
  text description of the shape.
\end{itemize}

\subsubsection{High resolution model v0}
\begin{figure}[htbp]
    \centering
    \includegraphics[width=7cm]{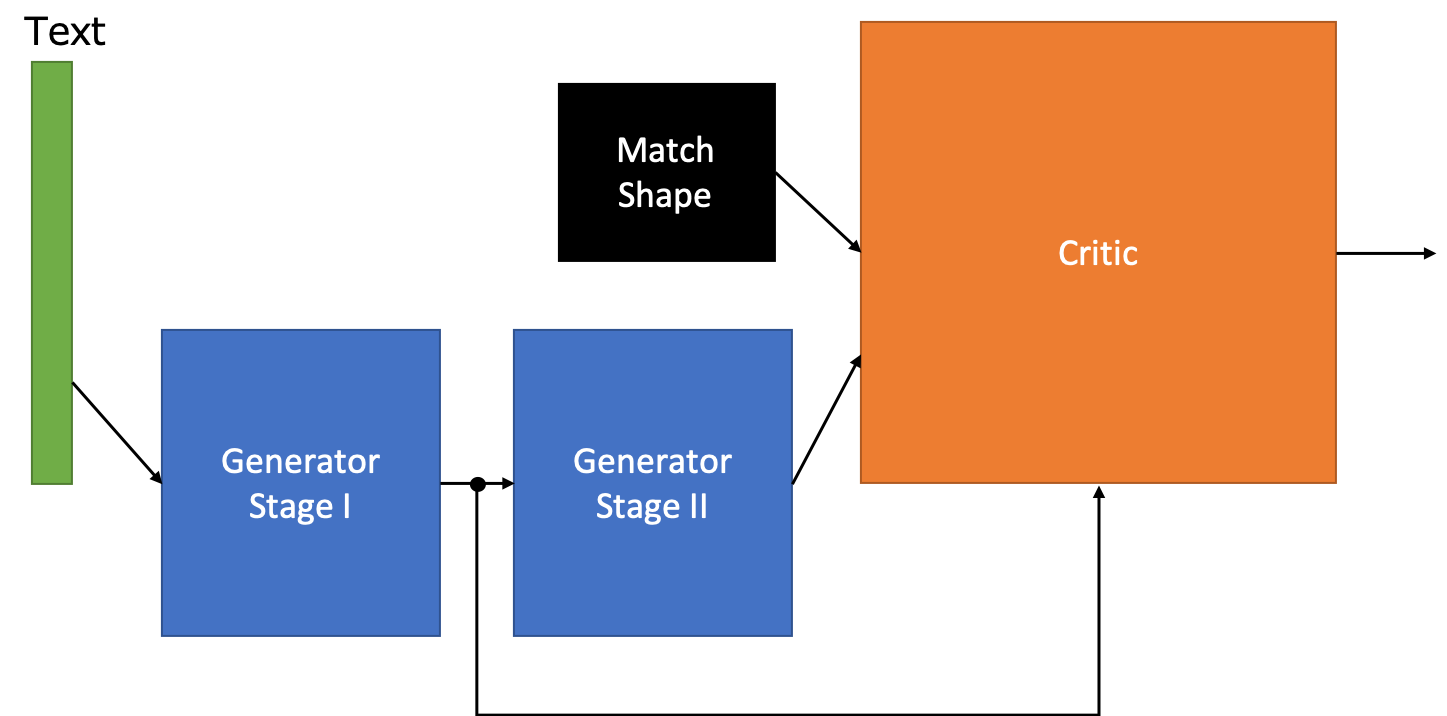}
    \caption{The flow of high resolution task v0}
    \label{v0flow}
\end{figure}

In this model, Critic monitors whether higher resolution shapes can be appropriately achieved from
the low resolution voxels. Fig.\ref{v0flow} shows the flow of v0 model.

A vector of text embedding is first input to the generator of StageI for generating low resolution voxels.
The output voxels are input to the generator of StageII to generate higher resolution voxels.
In v0 model, the input of the Critic network is both high resolution voxels and low resolution voxels so
that the Critic can evaluate high resolution voxels based on the information of low resolution ones.
We do not use vectors of text embedding for tasks after StageII because StageII concentrates
solely on heightening resolution. Therefore, high resolution tasks can be completely separated
from low resolution tasks by not using the text information in StageII.

\begin{figure}[htbp]
    \centering
    \includegraphics[width=7cm]{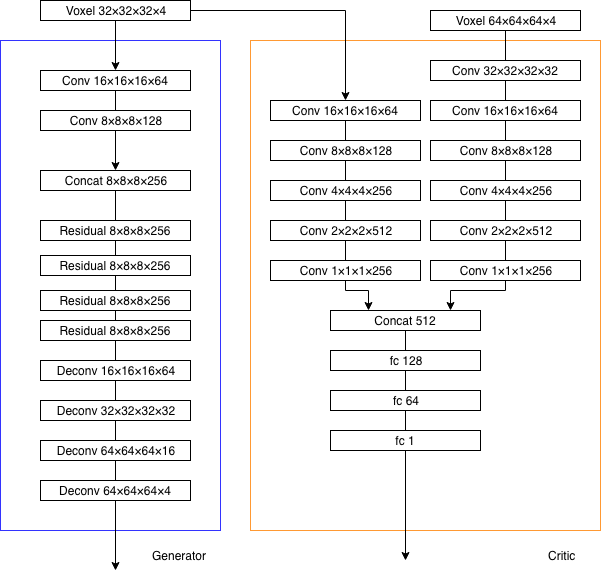}
    \caption{The model of high resolution task v0}
    \label{v0model}
\end{figure}

Fig.\ref{v0model} shows the training model of v0. We introduce a residual layer as a hidden layer of
the Generator network of StageII. By the residual layer, we can optimize each layer by learning the
residual function using layer input instead of learning the optimum output of each layer\cite{resnet}.

The loss function for v0 model is defined as follows:
\begin{align*}
    L_{CWGAN}&=\mathbb{E}_{t\sim p_\tau}[D(G_2(G_1(t)),G_1(t))]\\
    &-\mathbb{E}_{(t,s)\sim p_{mat}}[D(s,G_1(t))]+\lambda_{GP}L_{GP}\\
    L_{GP}&=\mathbb{E}_{(t,s)\sim p_{GP}}[(||\nabla_s D(s,G_1(t))||_2-1)^2\\
    &\qquad+(||\nabla_{G_1} D(s,G_1(t))||_2-1)^2]
\end{align*}
where the generators for low resolution and high resolution tasks are indicated by $G_1$ and
$G_2$, respectively. By relieving the term contributed to $ p_{mis} $ from $L_{CWGAN}$, the
model suppress learning the aspect of whether the shape matches with the text description.
The degree of matching between the text description and the generated shape depends on the
quality of training in StageI. However, by concentrating solely on heightening resolution, the
number of parameter updates in the training phase can be greatly reduced, resulting in faster
learning speed.

\subsubsection{High resolution model v1}
\begin{figure}[htbp]
    \centering
    \includegraphics[width=7cm]{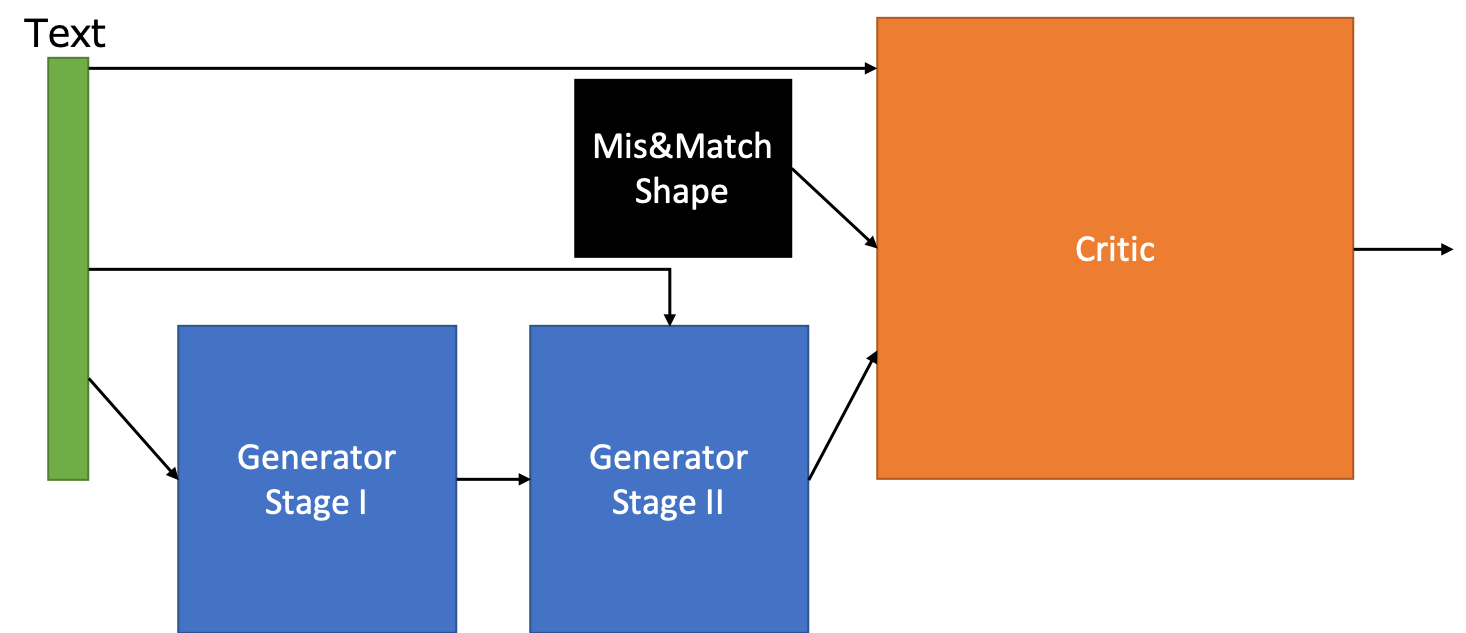}
    \caption{The flow of high resolution model v1}
    \label{v1flow}
\end{figure}

In this model, the Critic network monitors whether voxels are faithfully generated from the
text depictions as like TSGAN. Fig.\ref{v1flow} shows the flow of v1 model.

\begin{figure}[htbp]
    \centering
    \includegraphics[width=7cm]{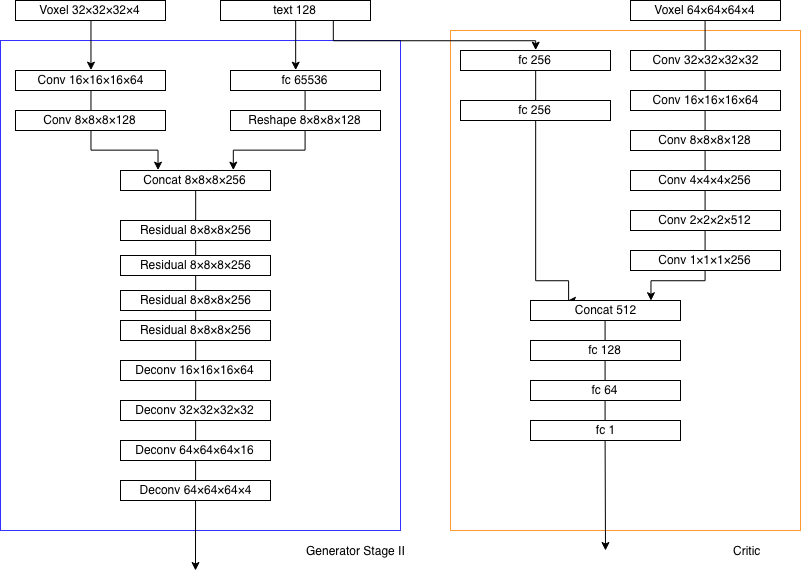}
    \caption{The training model of high resolution task v1}
    \label{v1model}
\end{figure}

A vector of text　embedding is first input to the generator of StageI for generating low resolution voxels.
Here, StageI is the same as Generator of Fig.\ref{stage1}. 
The generated low resolution voxel is input to StageII and a one-dimensional vector
is created through the convolution layers. The vector is combined with the vector of text embedding
and converted to a voxel again by the deconvolution layers. The reason of combining a vector representation of the voxel with the vector of text embedding is to generate a higher resolution voxel which reflects
the details described in the text. The Critic network tries to determines both how
realistic the generated voxel looks and how closely it matches with the text by using Wasserstein
distance. Fig.\ref{v1model} shows the training model of v1.

The loss function for v1 is defined as follows:
\begin{align*}
    L_{CWGAN}&=\mathbb{E}_{t\sim p_\tau}[D(t,G_2(G_1(t)))]+\mathbb{E}_{(t,s)\sim p_{mis}}[D(t,s)]\\
    &\quad-2\mathbb{E}_{(t,s)\sim p_{mat}}[D(t,s)]+\lambda_{GP}L_{GP}\\
    L_{GP}&=\mathbb{E}_{(t,s)\sim p_{GP}}[(||\nabla_t D(t,s)||_2-1)^2\\
    &\qquad+(||\nabla_s D(t,s)||_2-1)^2]
\end{align*}
This is the same loss function as TSGAN. Therefore, the Critic network evaluates whether the
generated voxel is properly created from the text. The cost for a voxel mismatched with the
text is included in the loss function to check how accurately the voxel is matched with the
text. However, many parameter updates are needed in training the network since the
Generator has two roles of heightening the resolution and making the generated voxel is closely
matched with the text. Therefore, long training time is expected in this model

\section{Evaluation}

\subsection{Experimental setup}


In this section, we evaluate the proposed high resolution tasks v0 and v1. We compare the
shapes generated by the models trained with 18000 epochs. In this evaluation, we use two
objects, table and chair for generating 3D shapes. For this experiment, we used the same
dataset as in TSGAN\cite{chen2018text2shape}. We created train/validation/test data by
randomly splitting the dataset into the ratio of $80\%/10\%/10\%$, respectively. As for text
embedding, we exclude texts which are not related to a table or a chair, texts with
misspelling, or text without enough description to generate voxels.

In order to evaluate the generated 3D shapes in the aspect of quality, we use the following two indices.
\begin{itemize}
\item Accuracy of classification (Class acc.): The first index is the accuracy rate
  of classification which evaluate how correctly the generated shapes are classified.
  We created another classifier network model to classify two target objects represented by
  voxels generated from the text descriptions. All the generated 3D voxels are input to
  the classifier and evaluate its accuracy rate. The model of the classifier is created
  based on the prior work \cite{VoxNet}. We expect that this accuracy metric can reflect
  how realistic the generated voxels look like.

\item Mean squared error (mse): The second index is the mean square error of the text embedding
  vector. We created an encoder that generates a 128 dimensional vector from a resulted voxel.
  The dimension of the vector is the same as the text embedding vector. We train this encoder
  to minimize the mean squared error between the output vectors and original text embedding vectors.
  We input the resulting voxels into the encoder network and then calculate the mse between
  the result and the corresponding latent vectors. We expect that this can evaluate how well
  each voxel is matched with the text representation for it.
\end{itemize}

\subsection{Generation result and its quantitative evaluation}

\begin{figure}[htbp]
    \centering
    \includegraphics[width=7cm]{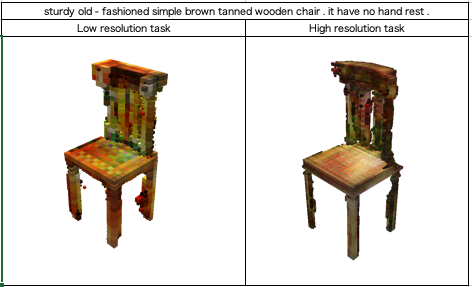}
    \caption{The example of generation in high resolution task v0 (18000 epochs)}
    \label{v018000}
\end{figure}

\begin{figure}[htbp]
    \centering
    \includegraphics[width=7cm]{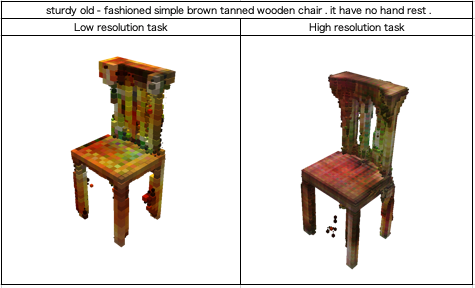}
    \caption{The example of generation in high resolution task v1 (18000 epochs)}
    \label{v118000}
\end{figure}

Fig.~\ref{v018000} and Fig.~\ref{v118000} show examples of generated result with v0 and v1 models.

Table~\ref{hyouka} shows the evaluation results of two indices for v0 and v1 models.
The higher the classification accuracy rate, the better the quality of generated voxels.
As for the mse, the lower the better. From the table, we see that
the classification accuracy rate of v1 is larger than that of v0 and the mse of v1 is
smaller than that of v0.

\begin{figure}[htbp]
    \centering
    \includegraphics[width=7cm]{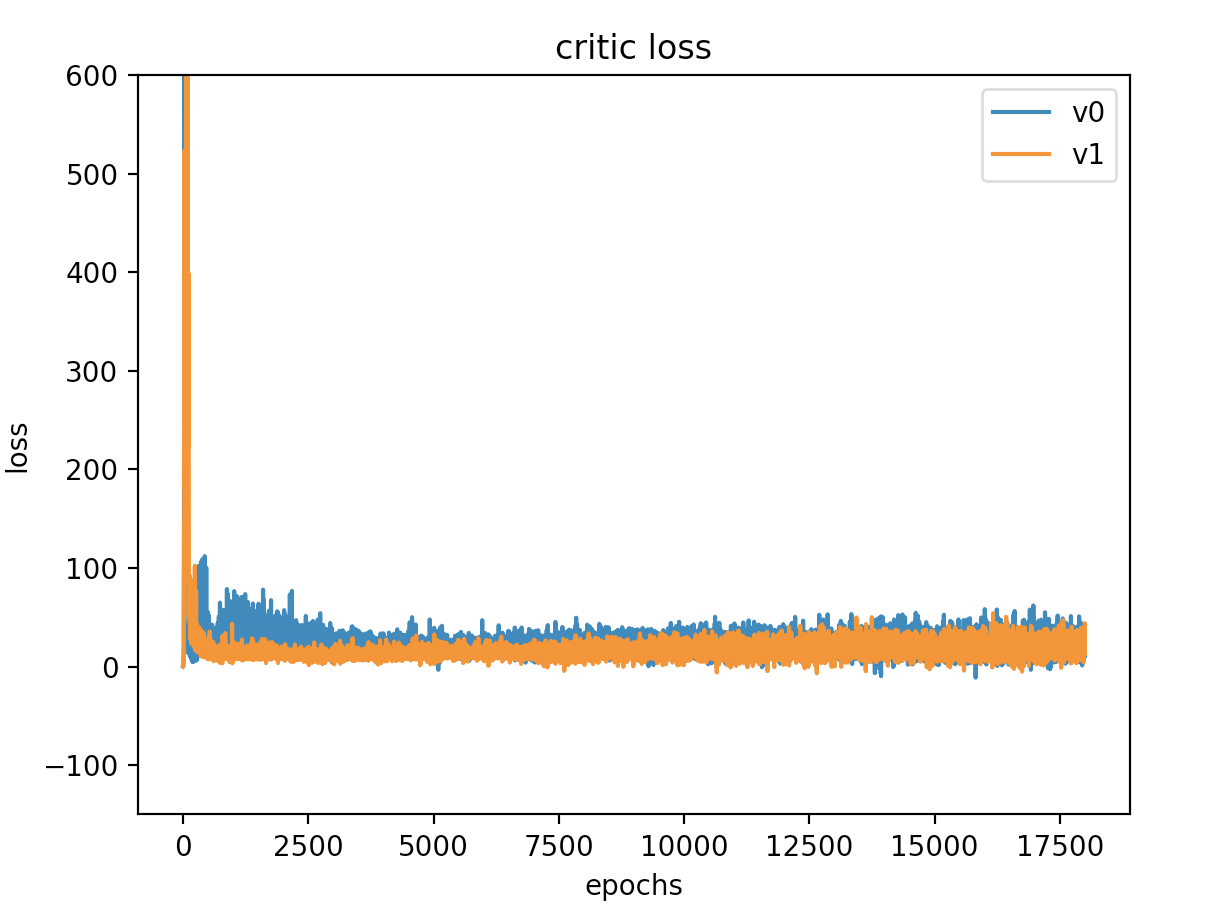}
    \caption{Critic's loss of high resolution task}
    \label{critic}
\end{figure}

\begin{figure}[htbp]
    \centering
    \includegraphics[width=7cm]{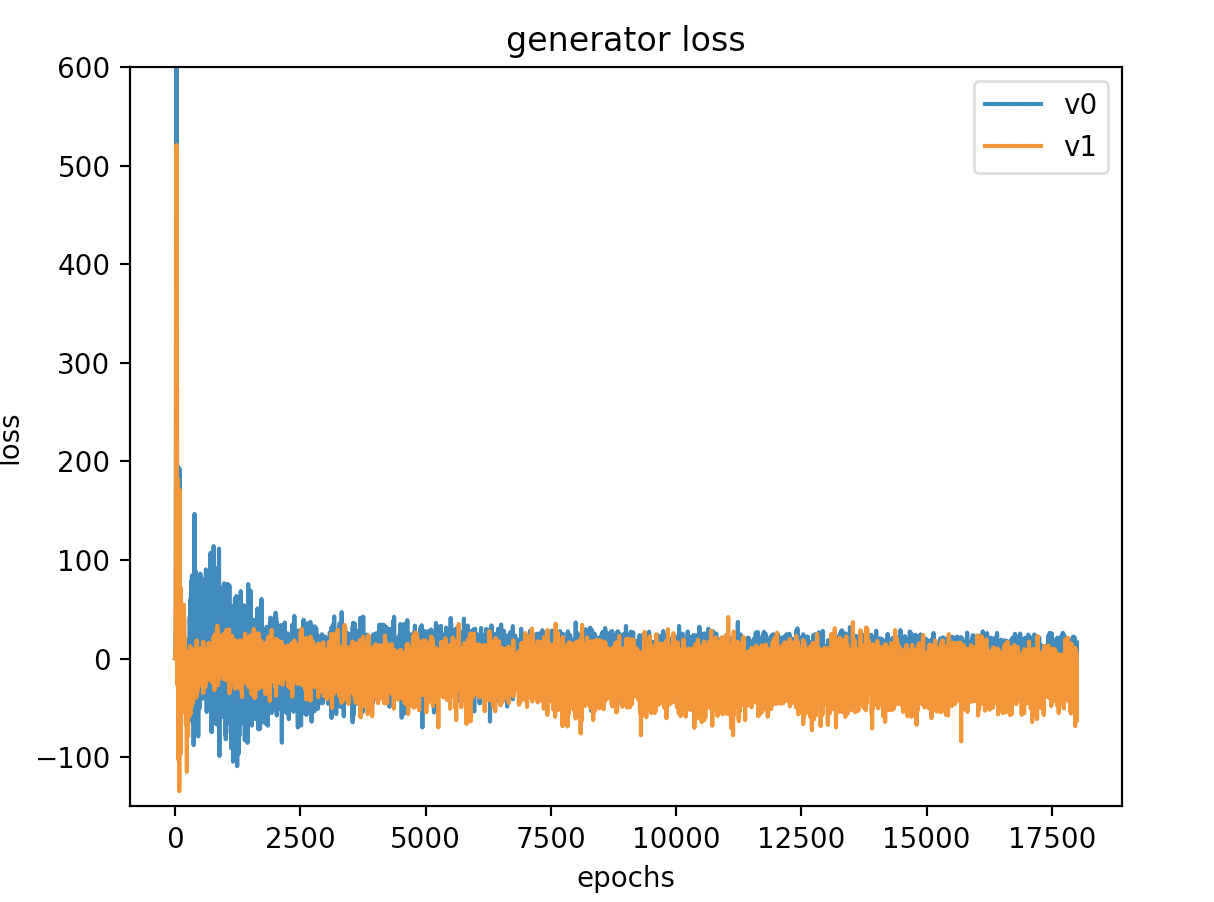}
    \caption{Generator's loss of high resolution task}
    \label{generator}
\end{figure}

Fig.~\ref{critic} and Fig~\ref{generator} show the trend of training losses of the Critic and
the Generator network for two high resolution tasks, respectively. The loss for the both the
overall trend of the loss for v0 and v1 in the Critic network is similar. However, variance of
the loss trend for v0 is larger than that of v1 in the Generator network.

\begin{table}[htbp]
    \caption{Evaluation of each model in two indices}
    \centering
    \label{hyouka}
    \begin{tabular}{ccc}
    \hline
    Method  & Class acc.  & mse  \\ \hline\hline
    DataSet & 1.0         & 0.153 \\
    v0      & 0.97        & 0.156 \\
    v1      & \bf{0.98}         & \bf{0.141} \\ \hline
    \end{tabular}
\end{table}

\subsection{Discussion}

As stated above, we compare the results of the networks trained with 18000 epochs.
At this epoch, both v0 and v1 models have almost no 3D shape collapse. According to
Table~\ref{hyouka}, v1 model is more faithful to the text than v0 model.
As for the accuracy, both v0 and v1 model have high classification accuracy, but the v1 model
achieves a little higher accuracy rate compared with the v0 model.

This is because the Critic in v1 model can calculate the distance of the probability distribution between
the train data and the generated voxel more accurately by referring to the text description.
Thus, the Generator could generate more realistic voxels properly.
The mean squared error for v1 is smaller that v0, meaning that v1 achieves better
encoding results. This indicates the v1 model can generate voxels more faithfully to the text.
Since we use information of the cost for a voxel mismatched with the text, the Critic can
calculate Wasserstein distance more properly.

As can be seen from Fig.~\ref{v018000} and Fig.~\ref{v118000} (and also the figures in
Appendix, the shape resolution is appropriately increased in both v0 and v1 models.
However, a little change in the color distribution can be seen compared with the low
resolution cases. We consider this is because the constraint of ``How realistic and faithful
the generated voxel is'' is applied to the loss function, but we do not set any constraint
regarding ``How a high resolution shape is faithful to the corresponding low resolution
shape''. As in the previous study\cite{StackGAN2}, introducing a constraint on the mean
and variance of the color distribution to the loss function may suppress the change in color.

\section{Conclusion}

In this paper, we extend the prior work \cite{chen2018text2shape} and propose new GAN models that
can generate high resolution voxels. In the proposed model, we also improved the previous
method to generate even low resolution voxels more faithfully to a given text.

The contributions of this research are three-fold. First, we proposed the models which generate
high resolution voxels faithfully to a given text from low resolution voxels. From the evaluation
results, it is possible to generate high resolution voxels with good visual quality. Second, we
contrived multiple roles of the Critic network and configured multiple models. We showed that
there was a difference in accuracy depending on whether separating the higher resolution task
from considering the text latent vector. Third, we introduced multiple indices to compare the
performance of the models.  As described in the discussion, there is a possibility for our
proposed model to generate voxels that are more faithful to a given text with a higher quality
shape. 
\bibliography{bibt}
\appendix
\section{Examples of Generation}
We attach the result of the voxel generated by the model of this experiment. All voxels are generated by model v1.
\begin{figure}[htbp]
    \centering
    \includegraphics[width=7cm]{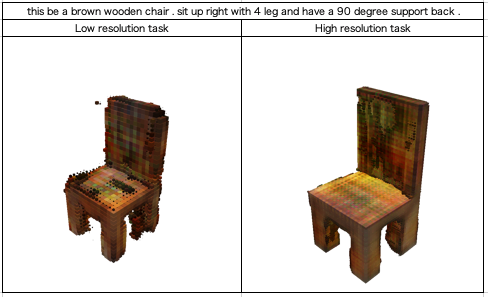}
    \label{1}
\end{figure}
\begin{figure}[htbp]
    \centering
    \includegraphics[width=7cm]{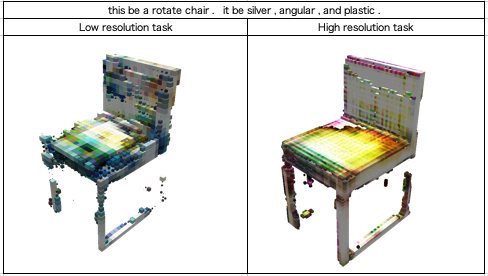}
    \label{2}
\end{figure}
\begin{figure}[htbp]
    \centering
    \includegraphics[width=7cm]{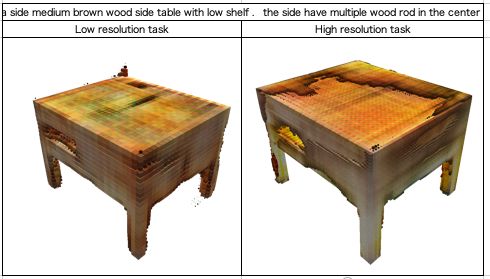}
    \label{3}
\end{figure}
\begin{figure}[htbp]
    \centering
    \includegraphics[width=7cm]{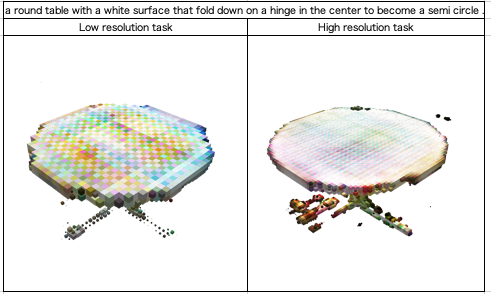}
    \label{4}
\end{figure}\begin{figure}[htbp]
    \centering
    \includegraphics[width=7cm]{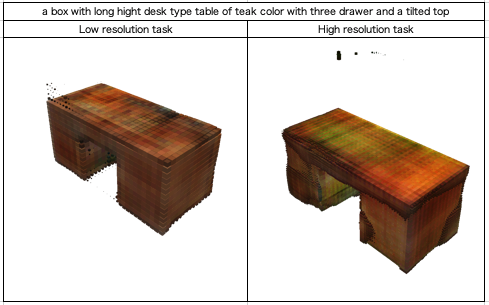}
    \label{5}
\end{figure}
\begin{figure}[htbp]
    \centering
    \includegraphics[width=7cm]{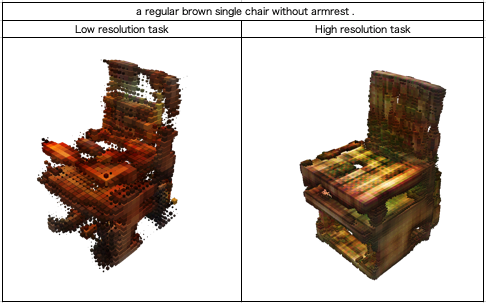}
    \label{6}
\end{figure}
\begin{figure}[htbp]
    \centering
    \includegraphics[width=7cm]{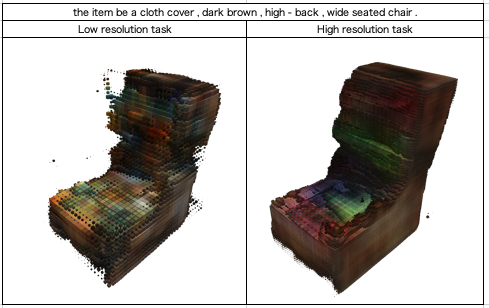}
    \label{7}
\end{figure}
\begin{figure}[htbp]
    \centering
    \includegraphics[width=7cm]{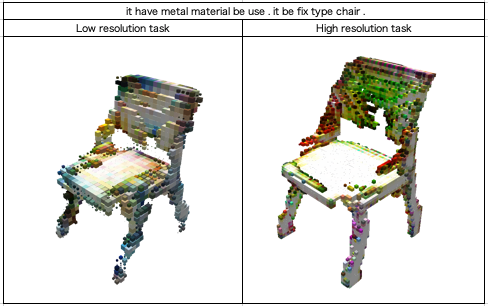}
    \label{8}
\end{figure}
\end{document}